\documentclass[aps,twocolumn,showpacs,amssymb,%
floatfix,superscriptaddress]{revtex4}
\usepackage{graphicx}
\usepackage{epsf}
\usepackage{dcolumn}
\usepackage{bm}
\usepackage{hyperref}
\usepackage{latexsym}
\usepackage{color}
\usepackage[croatian,english]{babel}
\usepackage{amssymb}
\usepackage{bm}
\usepackage{amsmath}

\newcommand{\beq}{\begin{equation}}
\newcommand{\eeq}{\end{equation}}

\newcommand{\bsy}{\boldsymbol}
\newcommand{\rarrow}{\rightarrow}
\newcommand{\arrows}{\leftrightarrow}
\begin{document}
\def\av#1{\langle#1\rangle}
\def\etal{{\it et al\/.}}
\def\pc{p_{\rm c}}
\def\l{{\lambda}}
\def\hm{h_*}
\def\xm{x_*}
\def\remark#1{{\bf *** #1 ***}}


\title{Influence of reciprocal arcs on the degree distribution and degree correlations}

\author{Vinko Zlati\'{c}\thanks{vzlatic@irb.hr}}
\affiliation{Theoretical Physics Division, Rudjer Bo\v{s}kovi\'{c} Institute, P.O.Box 180, HR-10002 Zagreb, Croatia}
\affiliation{INFM-CNR Centro SMC Dipartimento di Fisica, 
Sapienza Universit\`a di Roma Piazzale Moro 5, 00185 Roma, Italy }
\author{Hrvoje \v{S}tefan\v{c}i\'{c}\thanks{shrvoje@thphys.irb.hr}}
\affiliation{Theoretical Physics Division, Rudjer Bo\v{s}kovi\'{c} Institute, P.O.Box 180, HR-10002 Zagreb, Croatia}

\begin{abstract}
Reciprocal arcs represent the lowest order cycle possible to find in directed graphs without selfloops. Representing also a measure of feed-back between vertices, it is interesting to understand how reciprocal arcs influence other properties of complex networks. In this paper we focus on influence of reciprocal arcs on vertex degree distribution and degree correlations. We show that there is a fundamental difference between properties observed on the static network compared to the properties of networks which are obtained by simple evolution mechanism driven by reciprocity. We also present a way to statistically infer the portion of reciprocal arcs which can be explained as a consequence of feed-back process on the static network. In the rest of the paper the influence of reciprocal arcs on a model of growing network is also presented. It is shown that our model of growing network nicely interpolates between BA model for undirected and the BA model for directed networks.
\end{abstract}
\maketitle

\section{Introduction}

Most of real networks as for example Internet~\cite{CMP01}, WWW~\cite{AJB01} 
or biological webs~\cite{GCP03}, etc. show interesting topological properties compared with simple models of random graphs.
 Today it is well known that most of the real networks exhibit certain properties such as fat tail degree distributions or small-world effect etc.~\cite{N03, Boccaleti}.
 Reading literature on different types of networks a reader will find a huge number of papers describing different types of correlations in complex networks. 
Assortativity~\cite{Newman02f}, clustering coefficient~\cite{WS98}, reciprocity~\cite{Reciprocity}, k-cores~\cite{core}, rich-club coefficient~\cite{col}, Triad Significance Profile~\cite{Milo04} and many other measures related to correlations are frequently reported.
 Although the identification of these correlations and their reporting in various empirical complex networks has significantly improved our understanding of the field, the question of interrelations of correlation measures naturally emerges. 
Is it really surprising to find, for example, both the strong rich-club behavior and strong degree correlations in the network? The answer is -- clearly not. Today, it is a well known fact that most of the real networks are correlated.
 Nevertheless, there is still a huge gap in our understanding of how exactly certain types of correlation-related measures influence other correlation-related measures.
 In this paper we will try to bridge a part of that gap relating the reciprocity measure to degree sequence and to degree correlations.

The directed network represents an interesteng subgroup of real networks which allow movement in just one direction.
 Reciprocity~\cite{Reciprocity} of complex networks is fraction of arcs which have their counterparts showing in the opposite direction compared to the total number of arcs i.e. every bidirectional arrow in the directed graph is considered as composed of two reciprocal arcs.
 It can be said that it is in fact a measure of how much is directed network similar to undirected one.
 Reciprocity was also shown as an important feature for percolation on directed networks~\cite{Boguna}.
 In previous work we have also shown that the reciprocity is a very stable correlation measure of all the investigated measures in the case of Wikipedia networks ensemble~\cite{Zlatic_Wiki_2006}.

In the paper~\cite{RecipMoj} the influence of the broad class of degree correlations on the reciprocity measure is described and quantified.
 In this paper our aim is exactly the opposite i.e. to find a way to quantify the influence of reciprocal arcs on degree correlations and degree distributions of complex networks.
 More precisely, first we focus on random addition of reciprocal arcs in the underlying static network.
 This process of transformation of unidirectional arcs into the bidirectional ones can be justified in many ways.
 First, it is the simplest possible choice of creating reciprocal links in the network which already has some structure.
 Second, it is the logical model of information return in the case of the information networks like an e-mail or WWW network. 

\section{Influence of reciprocity on degree correlations in static networks}\label{SekcijaRecipCorrs}

As a null hypothesis it is reasonable to suppose that the mutual functional relationships between vertices are distributed completely randomly over the whole network.
 In other words, that means that we suppose that the mutual functional relationship does not depend in any way on the degrees of vertices or any other measurable network quantity.
 We suppose that reciprocal arcs can form only between vertices which are already connected. 
The question we address is the following: how reciprocal arcs, formed in this way, transform the degree distribution and correlations between degrees in a given complex network?
The model is defined with the initial network as an input.
 On the starting network the unidirectional arcs are transformed with probability $p$ into reciprocal ones, and with probability $1-p$ are left unchanged. 
After this process the properties of the new network are measured again. We show that using the inversion of the transformation process we can infer the most probable starting configuration of the network. 

In the following we distinguish between bidirectional arc as a single arc which is pointing in two directions contrary to some other analyses in which reciprocal arcs are represented as two arcs connecting two vertices $i$ and $j$ in opposing directions~\cite{Reciprocity, RecipMoj}.
 Every vertex of the network is described by $3$ numbers -- the first represents exclusive in-degree of the vertex, the second represents exclusive out-degree of the vertex, and the third represents its bidirectional degree.
 Exclusive in (or out) degree is the number of unidirectional in (or out) arcs which are attached to a given vertex.
 In the following these degrees will be designated as $\bsy{k}\equiv(k_i,k_o,k_r)$.  

From the initial network we can extract the following information as initial conditions for the observed transformation process: $L$ - number of arcs; $N$ - number of vertices; $L^\rarrow$ - number of \emph{strictly} unidirectional arcs; $L^\arrows$ - number of \emph{strictly} bidirectional arcs; $L(\bsy{k}\rarrow\bsy{q})$ - number of unidirectional arcs which are pointing from the vertex of degrees $\bsy{k}$ to the vertex of degrees $\bsy{q}$; $L(\bsy{k}\arrows\bsy{q})$ - number of bidirectional arcs which are connecting the vertices of degrees $\bsy{k}$ and degrees $\bsy{q}$; $N(\bsy{k})$ - number of vertices with degrees $\bsy{k}$. In the following text the convention will be that if we observe unidirectional arcs, with $\bsy{k}$ are designated degrees of the starting vertex while with $\bsy{q}$ are designated the degrees of the ending vertex.

With these properties it is possible to represent adequately maximally random graph, as well as graphs with any given degree distribution and correlations between degrees of neighboring vertices. 
Information on correlations between degrees of neighboring vertices existing in the network is given by frequency of arcs which connect different vertices: 

\begin{eqnarray}\label{statistics}
 \mathcal{P}(\bsy{k},\rarrow,\bsy{q})&=&\frac{L(\bsy{k}\rarrow\bsy{q})}{L},\nonumber\\
\mathcal{P}(\bsy{k},\arrows,\bsy{q})&=&\frac{L(\bsy{k}\arrows\bsy{q})}{L}.
\end{eqnarray}

Probabilities $\mathcal{P}(\bsy{k},\rarrow,\bsy{q})$ and $\mathcal{P}(\bsy{k},\arrows,\bsy{q})$ are defined as joint probabilities that the vertex of degrees $\bsy{k}$ is \emph{pointing to}/\emph{is connected to} the vertex of degrees $\bsy{q}$, with unidirectional arc in the former and with the bidirectional arc in the later case. The proper summation of this joint probabilities is
\begin{equation}\label{summation}
\sum_{\substack{\bsy{q}=\stackrel{\leftrightarrow}{\bsy{q}}_{0}\\\bsy{k}\geq\bsy{q}}}^{\infty}\mathcal{P}(\bsy{k},\arrows,\bsy{q})+\sum_{\substack{\bsy{k}=\stackrel{\rightarrow}{\bsy{k}}_{0}\\ \bsy{q}=\stackrel{\rightarrow}{\bsy{q}}_{0}}}^{\infty}\mathcal{P}(\bsy{k},\rarrow,\bsy{q})=1, 
\end{equation}
where $\stackrel{\leftrightarrow}{\bsy{q}}_{0}=(0,0,1)$, $\stackrel{\rightarrow}{\bsy{k}}_{0}=(0,1,0)$ and $\stackrel{\rightarrow}{\bsy{q}}_{0}=(1,0,0)$.
The summations are different for the bidirectional arcs compared with unidirectional arcs because $\mathcal{P}(\bsy{k},\arrows,\bsy{q})=\mathcal{P}(\bsy{q},\arrows,\bsy{k})=\frac{L(\bsy{k}\arrows\bsy{q})}{L}$.
Although the statistics of degrees of neighboring vertices gives relevant information on correlation structure of the given network, and the one-vertex statistics can in principle be easily calculated from that information, from analytical aspect we will show that is much easier to explicitly calculate one vertex degree correlations described with
\begin{equation}\label{1nStatistics}
P(\bsy{k})=\frac{N(\bsy{k})}{N}, 
\end{equation}
where $P(\bsy{k})$ represents the joint probability that the vertex has degrees $k_i$, $k_o$ i $k_r$.

In the studied model every unidirectional arc is transformed in a bidirectional one with the probability $p$.
 The equation which expresses a new joint probability that a vertex of degrees $\bsy{k}'$ is pointing to a vertex of degrees $\bsy{q}'$ via unidirectional arc is:
\begin{equation}\label{Bayes}
\mathcal{P}'(\bsy{k}',\rarrow,\bsy{q}')=\sum_{\mathcal{C}}\mathcal{T}(\bsy{k}',\rarrow,\bsy{q}'|\bsy{k},\rarrow,\bsy{q})\mathcal{P}(\bsy{k},\rarrow,\bsy{q}),
 \end{equation}
where $\mathcal{T}$ represents the transition probability for the given process. A prime on the probabilities means that they are calculated after the transformation process, while the absence of a prime means that the probabilities are calculated from the given starting network. The summation is run over the set $\mathcal{C}$ of unidirectional arcs which fulfill the following conditions: \textit{(i)} The number of neighbors $S^{(j)}=k_i^{(j)}+k_o^{(j)}+k_r^{(j)}$ is conserved for every vertex $j$, because transformation process does not create new arcs between vertices which are not neighbors already; \textit{(ii)} Before and after the transformation process the following relations hold: $k'^{(j)}_i\leq k_i^{(j)}$, $k'^{(j)}_o\leq k_o^{(j)}$ and $k'^{(j)}_r\geq k_r^{(j)}$. The transition probability $\mathcal{T}$ written in a more detail is:
\begin{eqnarray}\label{T}
\mathcal{T}(\bsy{k}',\rarrow,\bsy{q}'|\bsy{k},\rarrow,\bsy{q})&=&(1-p)\mathcal{T}(k'_i|k_i)\mathcal{T}(k'_o-1|k_o-1)\nonumber\\
&&\mathcal{T}(q'_i-1|q_i-1)\mathcal{T}(q'_o|q_o).
\end{eqnarray}
\indent The first part of the equation (\ref{T}) is probability that the unidirectional arc stays unidirectional after the transformation process. Other unidirectional arcs attached to the vertices can be changed with probability $p$ or stay unidirectional with probability $1-p$. The fact that in this case only other arcs are monitored is represented in equation by subtracting one arc from the out-degree of the out vertex and the in-degree of the in vertex. Probabilities of the transition $\mathcal{T}(x'|x)$, where $x$ represents any of the aforementioned degrees are binomial probabilities i.e.
\beq\label{Binom}
\mathcal{T}(x'|x)=\binom{x}{x'}p^{x-x'}(1-p)^{x'}.
\eeq
\indent New joint probability distribution of degrees of the vertices connected via the bidirectional arc  $\mathcal{P}'(\bsy{k},\arrows,\bsy{q})$ is:
\begin{widetext}
\begin{eqnarray}\label{R'}
 \mathcal{P}'(\bsy{k}',\arrows,\bsy{q}')&=&\sum_{\mathcal{C}}\left(\mathcal{T}(\bsy{k}',\arrows,\bsy{q}'|\bsy{k},\arrows,\bsy{q})\mathcal{P}(\bsy{k},\arrows,\bsy{q})
+\mathcal{T}(\bsy{k}',\arrows,\bsy{q}'|\bsy{k},\rarrow,\bsy{q})\mathcal{P}(\bsy{k},\rarrow,\bsy{q})\right)\nonumber\\
&&+\sum_{\mathcal{C}'}\mathcal{T}(\bsy{q}',\arrows,\bsy{k}'|\bsy{k},\rarrow,\bsy{q})\mathcal{P}(\bsy{k},\rarrow,\bsy{q}).
\end{eqnarray}
\end{widetext}
\indent\indent The set $\mathcal{C}$ is fulfilling all the conditions as in equation (\ref{Bayes}), while for the set $\mathcal{C}'$ the following relations hold: \textit{(i)} The number of neighbors $S^{(j)}=k_i^{(j)}+k_o^{(j)}+k_r^{(j)}$ is conserved for every vertex $j$, because transformation process does not create new arcs between vertices which are not neighbors already; \textit{(ii)} Before and after the transformation process the following relations hold: $q'^{(j)}_i\leq k_i^{(j)}$, $q'^{(j)}_o\leq k_o^{(j)}$ and $q'^{(j)}_r\geq k_r^{(j)}$. In this equation the probabilities of transition have the similar meaning as in the equation (\ref{T})
\begin{eqnarray}\label{T1}
 \mathcal{T}(\bsy{k}',\arrows,\bsy{q}'|\bsy{k},\arrows,\bsy{q})&=&\mathcal{T}(k'_i|k_i)\mathcal{T}(k'_o|k_o)\nonumber\\
&&\mathcal{T}(q'_i|q_i)\mathcal{T}(q'_o|q_o),
\end{eqnarray}
while
\begin{eqnarray}\label{T2}
 \mathcal{T}(\bsy{k}',\arrows,\bsy{q}'|\bsy{k},\rightarrow,\bsy{q})&=&p\mathcal{T}(k'_i|k_i)\mathcal{T}(k'_o|k_o-1)\nonumber\\
&&\mathcal{T}(q'_i|q_i-1)\mathcal{T}(q'_o|q_o),
\end{eqnarray}
and
\begin{eqnarray}\label{T3}
 \mathcal{T}(\bsy{q}',\arrows,\bsy{k}'|\bsy{k},\rightarrow,\bsy{q})&=&p\mathcal{T}(k'_o|q_o)\mathcal{T}(k'_i|q_i-1)\nonumber\\
&&\mathcal{T}(q'_o|k_o-1)\mathcal{T}(q'_i|k_i).
\end{eqnarray}
\indent It is important to notice that in the equation (\ref{T1}) we do not need to worry about the arc which connects neighboring vertices because it is, process invariant, bidirectional arc. Parameter $p$ in equations (\ref{T2}) and (\ref{T3}) represents the probability of the transformation of unidirectional arc which connects neighbors to bidirectional arc. Individual probabilities of transition in equations (\ref{T1}), (\ref{T2}), and (\ref{T3}) are again binomial (\ref{Binom}).

Similar equations are easy to write for the transformation process of one vertex statistics. More precisely, equation
\begin{equation}\label{1node Trans}
P'(\bsy{k}')=\sum_{\mathcal{C}}\binom{k_i}{k'_i}\binom{k_o}{k'_o}p^{k_i-k'_i+k_o-k'_o}(1-p)^{k'_i+k'_o}P(\bsy{k}),
\end{equation}
describes their probability of transformation of joint probability of one vertex degrees described with $P(\bsy{k})$ into the joint probability of one vertex degrees $P'(\bsy{k}')$.

In all the aforementioned equations for the transformation process we used the joint probability statistics, because the total number of arcs over which this statistics is obtained does not change with the process. However to calculate the correlations existing in the network it is convenient to use the conditional probabilities with respect to the type of the arc which connects two neighboring vertices. The usual equation for conditional probability can be employed as
\beq\label{JointToCondUni}
\mathcal{P}'(\bsy{k}',\bsy{q}'|\rarrow)=\mathcal{P}'(\bsy{k}',\rarrow,\bsy{q}')/P'(\rarrow).
\eeq
 The probability that two neighboring vertices are connected with unidirectional arc is $P'(\rarrow)=P(\rarrow)(1-p)$ and the probability that an unidirectional arc exists before the transformation process is $P(\rarrow)=L^{\rarrow}/L$, where $L^{\rarrow}$ is the number of unidirectional arcs before the transformation process.
 Similar equation holds for conditional probability that two vertices are connected via bidirectional arc:
\beq\label{JointToCondBi}
\mathcal{P}'(\bsy{k}',\bsy{q}'|\arrows)=\mathcal{P}'(\bsy{k}',\arrows,\bsy{q}')/P'(\arrows),
\eeq
where  $P'(\arrows)=P(\arrows)+pP(\rarrow)$ and $P(\arrows)=L^{\arrows}/L$. The $L^{\arrows}$ is the number of bidirectional arcs before the transformation process.
\begin{table}[t]
\centering
\begin{small}
\begin{tabular}{|l|l|}
\hline
$\rarrow$ & $\langle k_\mu k_\nu|\rarrow\rangle$ \\ 
statistics
 & $\langle k_\mu q_\nu|\rarrow\rangle$ \\ 
$\mu,\nu\in\{i,o,r\}$ & $\langle k_\mu |\rarrow\rangle$ \\ 
&$\langle q_{\mu}|\rarrow\rangle$ \\
\hline
$\arrows$ & $\langle k_\mu k_\nu|\arrows\rangle$ \\ 
statistics & $\langle k_\mu q_\nu|\arrows\rangle$ \\ 
$\mu,\nu\in\{i,o,r\}$ & $\langle k_\mu |\arrows\rangle$ \\
\hline 
1-vertex &  $\langle k_\mu k_\nu\rangle$\\ 
$\mu,\nu\in\{i,o,r\}$
 &  $\langle k_\mu \rangle$\\ 
\hline
\end{tabular}
\end{small}
\caption{Table of studied statistical moments and correlations. In the first column is designated if the averaging is performed over vertices, unidirectional or bidirectional arcs. The second column describes interesting product moments.}
\label{TablicaProucavanja}
\vspace{-0.5cm}
\end{table}
\begin{figure}[t]
\centering
\includegraphics*[width=0.4\textwidth]{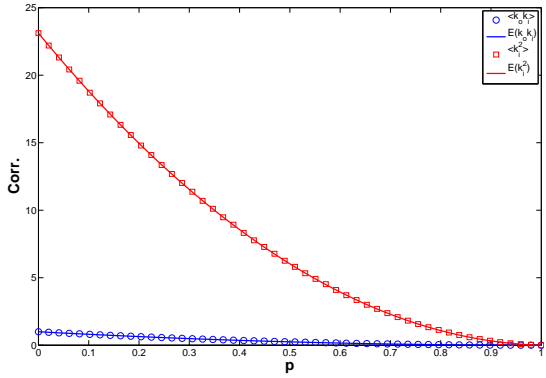}
\caption[Promjena jednovr\v{s}nih korelacija sa parametrom $p$]{\label{SlikaPromKorel} This figure shows an excellent agreement between simulations of transformation process and our analytical treatment. On x-coordinate is parameter $p$ which represents the probability of arc transformation. The y-coordinate represents the numerical value of given correlations. Simulations are designated by markers and analytical results with lines. The initial network is a Barab\'asi-Albert directed network of $10^5$ vertices. The simulations are averaged over $1000$ realizations.}
\end{figure}
In the literature on complex networks \cite{N03} it is usual to use statistics of average degree of neighbors of a given vertex to represent the correlations of degrees in the network. 
Such a measure is usually represented with figures of the average neighboring degree dependence on the degree of the monitored vertex.
It is easy to verify if the network is correlated or not by simple inspection of such a figure. 
In order to analytically describe degree--degree correlations resulting from this process we will use a different measure much more common in usual statistical analysis. 
The observed and calculated correlations are just the noncentralized product-moments of independent variables. 
In the following we will loosely use the term correlations for all of the calculated statistical quantities both for statistics obtained on one vertex via Eq. (\ref{1node Trans}) or for statistics of degrees on connected pairs of vertices calculated via Eq. (\ref{Bayes}) and (\ref{R'}). 
All the possible types of correlations we studied are summarized in Table \ref{TablicaProucavanja}. 

The equation for calculation of 1-vertex statistics is:
\beq\label{OneVertexCals}
\langle k'_ik'_o\rangle=\sum_{\bsy{k'}}k'_ik'_o P'(\bsy{k}'),
\eeq
for the case of in-out degree correlations. 
All other 1-vertex degree correlations are calculated in a similar way. 
There are two different equations with which we calculate  2-vertex degree correlations. 
The first one is for the calculation of 2-vertex degree correlations connected via unidirectional arcs.
 This type of correlations are designated as $\langle \cdot|\rarrow\rangle$ in order to distinguish them from the 2-vertex degree correlations calculated via bidirectional arcs $\langle \cdot|\arrows\rangle$. 
The equation for the in-out degree correlations of the vertices connected via unidirectional arc is: 
\beq\label{kiko}
\langle k'_iq'_o|\rarrow\rangle=\sum_{\bsy{k'}\bsy{q'}}k'_iq'_o\mathcal{P}'(\bsy{k}',\bsy{q}'|\rarrow).
\eeq
Using equations (\ref{Bayes}) and (\ref{JointToCondUni}) we can calculate degree correlations of unidirectionally connected pairs of vertices. 
\begin{figure}[t]
\centering
\includegraphics*[width=0.4\textwidth]{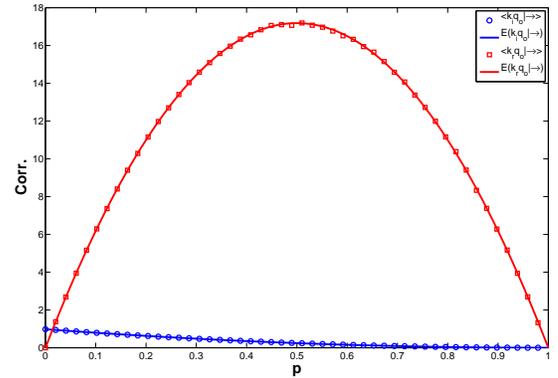}
\caption[Promjena dvovr\v{s}nih korelacija sa parametrom $p$]{\label{Aver2NodeUnidir.eps} The evolution of 2-vertex correlations, measured on pairs connected with unidirectional arcs, with change of parameter $p$. The initial network is a unidirectional version of Barab\'{a}si-Albert model of size $10^5$ with average out degree equal to one. We averaged over $100$ different realizations of the process on $10$ different realizations of directed B-A model. The measured correlations are shown in Figure as markers.  Analytical results are  presented as full lines. Agreement between simulation and analytics is excellent. }
\end{figure}
\indent The calculation of degree correlations of bidirectionally connected pairs of vertices is a bit trickier. The approximate equation for in-out degree correlations in this case is:
\beq\label{kikoBi}
2\langle k'_iq'_o|\arrows\rangle\simeq\sum_{\bsy{k'}\bsy{q'}}k'_iq'_o\mathcal{P}'(\bsy{k}',\bsy{q}'|\arrows).
\eeq
This equation is just an approximation because in order to analytically calculate expected correlations after the transformation process we have to sum over all degrees $\bsy{k'}$ and $\bsy{q'}$ thus including every bidirectionally connected pair two times, except for the pairs which have \emph{exactly the same} degrees. The equation could be improved by taking into account a new class of correlations just between the bidirectionally connected pairs of vertices which have the same degrees, but as it will be shown later, this approximation is more than good enough for estimating expected correlations for most of large enough networks. Using equations (\ref{R'}) and (\ref{JointToCondBi}) it is possible to calculate degree correlations of bidirectionally connected pairs of vertices.
\begin{figure}[t]
\centering
\includegraphics*[width=0.5\textwidth]{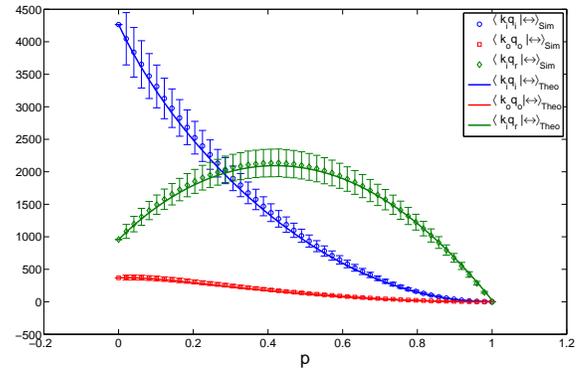}
\caption{\label{SimVsTheoRecip.eps} The change of some degree correlations of bidirectionally connected vertex pairs with parameter $p$. The initial network is the network of Spanish Wikipedia. We simulated $100$ different realizations of transformation process. There is a good agreement between the simulations (markers) and analytical results (full lines). Because analytical equation for these cases is approximate, the standard deviation margins are also plotted in the figure. The equation which describes the first type of correlations is Eq. (\ref{kiqiR}). The equation for second type of correlations is easily obtained with a change of indices. The equation for the third type of correlations is not presented in this paper, but can be obtained using the presented formalism.}
\end{figure}

\indent If the calculated correlations are less or greater than the expected in the random network of the same degree distribution (configuration model)~
\cite{Bekessy, Molloy}, then the network exhibits a structural tendency that the vertices of larger degrees are mutually connected less or more frequently.
 For example the expected in-out degree correlations of unidirectionally connected pairs of vertices is:
\beq\label{kikoRand}
\langle k'_iq'_o|\rarrow\rangle_{Rand}=\sum_{\bsy{k'}\bsy{q'}}k'_iq'_o\mathcal{P}'(\bsy{k}'|\rarrow)\mathcal{P}'(\bsy{q}'|\rarrow).
\eeq
In Eq. (\ref{kikoRand}) we use the conditional probabilities that the two neighboring vertices are connected via unidirectional arc.
 The $\mathcal{P}'(\bsy{k}'|\rarrow)$ designates that the vertex of degrees $\bsy{k}'$ is the starting vertex of the conditioned arc, while $\mathcal{P}'(\bsy{q}'|\rarrow)$ designates that the vertex with degrees $\bsy{q}'$ is the end vertex of that arc.
In the case of vertex statistics it can be written as
\beq\label{calPtoP}
\mathcal{P}'(\bsy{k}'|\rarrow)=\frac{k_o}{\langle k_o\rangle}P'(\bsy{k}'),
\eeq
because of the fact that the vertex certainly has outgoing arc which connects it to a neighboring vertex.

\indent It is important to understand the fine difference between the correlations between degrees of neighboring vertices and correlations of degrees of one vertex.
 In the following the correlations of degrees of neighboring vertices will be designated with the conditional type of arc, to make the distinction from one vertex correlations.
The studied types of correlations are shown in Table \ref{TablicaProucavanja}.
All the correlations are calculated analytically and checked with numerical simulations on networks of different sizes, density of arcs and starting correlation structure.
 In Figures \ref{SlikaPromKorel}, \ref{Aver2NodeUnidir.eps} and \ref{SimVsTheoRecip.eps} are shown some 1-vertex, 2-vertex unidirectional and 2-vertex bidirectional correlations calculated analytically and compared to simulations. 

For example the in -- in degree correlations of neighboring vertices can be calculated using the expression
\begin{equation}\label{kiqiFirstStep}
\langle k'_iq'_i|\rarrow\rangle=\sum_{\bsy{k}'\bsy{q}'}k'_iq'_i\mathcal{P}'(\bsy{k}',\bsy{q}'|\rightarrow),
\end{equation}
and using Eq. (\ref{Bayes}) and (\ref{JointToCondUni}) the final solution is
\begin{equation}\label{kiqiFinal}
\langle k'_iq'_i|\rarrow\rangle=(1-p)^2\langle k_iq_i|\rarrow\rangle+p(1-p)\langle k_i|\rarrow\rangle.
 \end{equation}
\indent The other example are in -- in degree correlations of bidirectioanlly connected vertices calculated with the presented scheme:
\begin{widetext}
\begin{equation}\label{kiqiR}
\langle k'_iq'_i|\arrows\rangle=\frac{(1-p)^2\left(\langle k_iq_i|\arrows\rangle P(\arrows)+pP(\rarrow)\left(\langle k_iq_i|\rarrow\rangle-\langle k_i|\rarrow\rangle\right)\right)}{P(\arrows)+pP(\rarrow)}.
\end{equation}
\end{widetext}
In this case the factors $P(\rarrow)$ and $P(\arrows)$ are also present because they did not cancel out as they did in Eq. (\ref{kiqiFinal}).

\indent To compute~(\ref{kiqiFirstStep}) and other possible correlations the following set of relations is useful:
\begin{eqnarray}\label{binomrelacije}
\sum_{l=0}^n \binom{n}{l}l^2\left(\frac{1-p}{p}\right)^l&=&\frac{n(1-p)}{p^n}((1-p)n+p)\nonumber\\
\sum_{l=0}^n \binom{n}{l}l\left(\frac{1-p}{p}\right)^l&=&\frac{n(1-p)}{p^n}\nonumber\\
\sum_{l=0}^n \binom{n}{l}\left(\frac{1-p}{p}\right)^l&=&\frac{1}{p^n}\nonumber\\
\sum_{l=1}^n \binom{n-1}{l-1}l\frac{(1-p)^{l-1}}{p^l}&=&\frac{n(1-p)+p}{p^n}\nonumber\\
\sum_{l=1}^n \binom{n-1}{l-1}\frac{(1-p)^{l-1}}{p^l}&=&\frac{1}{p^n}
\end{eqnarray}
\indent The computation of all elementary correlations shown in Table  \ref{TablicaProucavanja} can be more elegantly described with matrices of transformation $\mathbf{T}$. 
If the observed correlations are represented as components of a ``correlation vector'' the studied process can be described with two different transformation matrices - one which transforms vector of one vertex correlations $\mathbf{T}_{1v}$ and the second which transforms the vectors of the neighboring pairs correlations $\mathbf{T}_{2v}$.

\section{Transformation matrix}
\indent Two transformation matrices differ one from another. The matrix of one-vertex correlations is the square matrix of rank $8$. The complete one-vertex statistics of interest can be written as a vector $\mathbf{S}$ with components: $\mathbf{S^{\mathbf{T}}}=\left\{\langle k_i \rangle=\langle k_o \rangle,\langle k_r \rangle,\langle k^2_i \rangle,\langle k^2_o \rangle,\langle k^2_r \rangle,\langle k_ik_o \rangle,\langle k_ik_r \rangle,\langle k_ok_r \rangle\right\}$. The expected correlations calculated after the transformation process for one-vertex correlations can now be written as a simple linear equation
\beq\label{JednadzbaTransfromacije1v}
\langle \mathbf{S}'(p)\rangle=\mathbf{T}_{1v}(p)\mathbf{S}(0),
\eeq
where $\mathbf{S}'(p)$ represents vector of correlations after fraction $p$ of unidirectional arcs is transformed into bidirectional arcs. 
Average in-degree is always equal to average out-degree and is therefore eliminated from the matrix.
The transformation matrix for 1-vertex correlations $\mathbf{T}_{1v}$ is:
\begin{equation}\label{T_1v}
\mathbf{T}_{1v}=\left(
\begin{smallmatrix}
(1-p) & 0 & 0 & 0 & 0 & 0 & 0 & 0\\ 
2p & 1 & 0 & 0 & 0 & 0 & 0 & 0 \\
p(1-p) & 0 & (1-p)^2 & 0 & 0 & 0 & 0 & 0 \\
p(1-p) & 0 & 0 & (1-p)^2 & 0 & 0 & 0 & 0 \\
2p(1-p) & 0 & p^2 & p^2 & 1 & 2p^2 & 2p & 2p\\
0 & 0 & 0 & 0 & 0 & (1-p)^2 & 0 & 0 \\
-p(1-p) & 0 & p(1-p) & 0 & 0 & p(1-p) & (1-p) & 0 \\
-p(1-p) & 0 & 0 & p(1-p) & 0 & p(1-p) & 0 & (1-p) \\
\end{smallmatrix}\right).
\end{equation} 

\indent This matrix also has its inverse:
\begin{equation}\label{T-1_1v}
\mathbf{T}^{-1}_{1v}=\left(
\begin{smallmatrix}
\frac{1}{1-p} & 0 & 0 & 0 & 0 & 0 & 0 & 0\\ 
\frac{2p}{p-1} & 1 & 0 & 0 & 0 & 0 & 0 & 0 \\
-\frac{p}{(1-p)^2} & 0 & \frac{1}{(1-p)^2} & 0 & 0 & 0 & 0 & 0 \\
-\frac{p}{(1-p)^2} & 0 & 0 & \frac{1}{(1-p)^2} & 0 & 0 & 0 & 0 \\
-\frac{2p}{(1-p)^2} & 0 & \frac{p^2}{(1-p)^2} & \frac{p^2}{(1-p)^2} & 1 & \frac{2p^2}{(1-p)^2} & \frac{2p}{(1-p)} & \frac{2p}{(1-p)}\\
0 & 0 & 0 & 0 & 0 & \frac{1}{(1-p)^2} & 0 & 0 \\
\frac{p}{(1-p)^2} & 0 & -\frac{p}{(1-p)^2} & 0 & 0 & -\frac{p}{(1-p)^2} & \frac{1}{(1-p)} & 0 \\
\frac{p}{(1-p)^2} & 0 & 0 & -\frac{p}{(1-p)^2} & 0 & -\frac{p}{(1-p)^2} & 0 & \frac{1}{(1-p)} \\
\end{smallmatrix}\right).
\end{equation}
The inverse matrix can be of interest for statistical analysis of real networks. If there is a reason to believe that the bidirectional arcs are completely random consequence of the mentioned transformation process and if one has a network model which does not take into account the bidirectional arcs - it can be tested using the inverse transformation matrix. It is easy to calculate the parameter $p$ as $p=\frac{L^{\arrows}}{L}$, where $L^{\arrows}$ represents the number of bidirectional arcs in the network of interest, while $L$ is the total number of arcs. Then using equation
\beq\label{JednadzbaInvTransfromacije1v}
\langle \mathbf{S}(0)\rangle=\mathbf{T}_{1v}^{-1}(p)\mathbf{S}'(p),
\eeq
one can find the vector of expected degree correlations before the transformation process.
 Comparing then $\langle \mathbf{S}(0)\rangle$ with the vector of correlations obtained by the model one can gain additional information on the structural role of bidirectional arcs and/or quality of the studied model.
 Such assumptions could be a good null model for a number of real world applications such as analysis of communication or traffic networks. 
In the companion paper~\cite{RecipWiki} we will present application of this framework to the Wikipedia networks as a case study.
\begin{figure}[t]
\centering
\includegraphics*[width=0.45\textwidth]{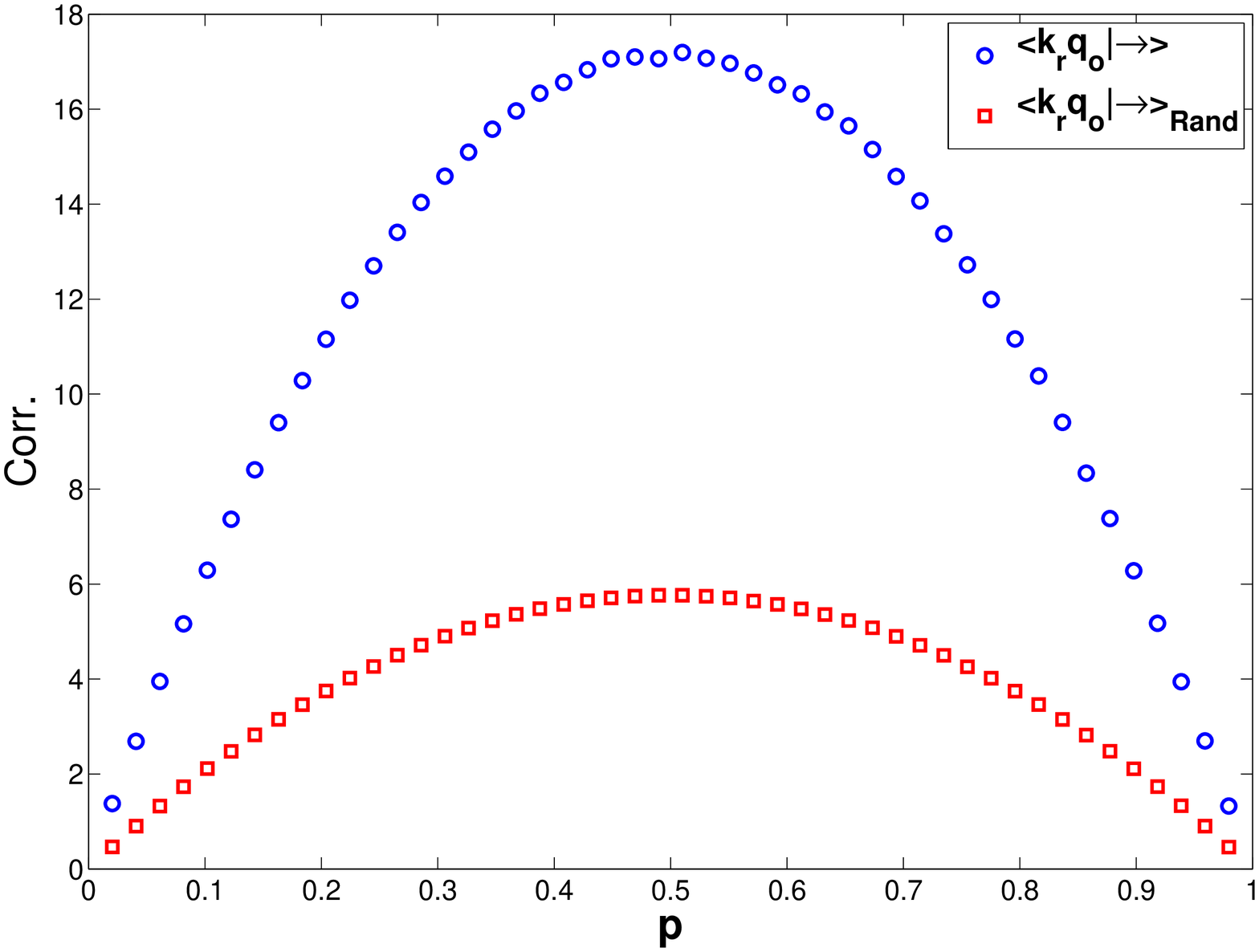}
\caption[Usporedba induciranih i o\v{c}ekivanih korelacija]{\label{CorrVsRandCorr.eps} The correlations $\langle k'_rq'_o|\rarrow\rangle$ resulting from the transformation process compared with the expected correlations of  configuration model $\langle k'_rq'_o|\rarrow\rangle_{Rand}= \frac{\langle k_rk_o\rangle\langle k_ik_o\rangle}{\langle k_o\rangle^{2}}$. In this case the transformation process clearly amplifies observed degree correlations. }
\end{figure}

The transformation of 2-vertex correlations is given by the expression
\begin{equation}\label{2vLinEq}
\langle \mathbf{S}_{2v}'(p)\rangle=\mathbf{T}_{2v}(p)\mathbf{S}_{2v}(0)+\mathbf{b}(p),
 \end{equation}
where $\mathbf{S}_{2v}$ presents vector of 2-vertex product moments and $\mathbf{b}(p)$ additional vector containig terms like $p^2$ given in Eq. (\ref{krqr}). 
The matrix of 2-vertex degree correlations is to big to be presented. For example, the equation for expected correlation of two bidirectional degrees of nodes connected via unidirectional arc is:
\begin{widetext}
\begin{eqnarray}\label{krqr}
\langle k'_rq'_r|\rarrow\rangle&=&\langle k_r,q_r|\rarrow\rangle-p\left(\langle q_r|\rarrow\rangle+\langle k_r|\rarrow\rangle\right)+p\left(\langle k_iq_r|\rarrow\rangle+\langle k_oq_r|\rarrow\rangle+\langle k_rq_o|\rarrow\rangle+\langle k_rq_i|\rarrow\rangle\right)\nonumber\\
&+&p^2\left(\langle k_iq_i|\rarrow\rangle+\langle k_iq_o|\rarrow\rangle+\langle k_oq_i|\rarrow\rangle+\langle k_oq_o|\rarrow\rangle-\langle k_i|\rarrow\rangle-\langle k_o|\rarrow\rangle-\langle q_o|\rarrow\rangle-\langle q_i|\rarrow\rangle+1\right).
\end{eqnarray}
\end{widetext}
Nevertheless, there is enough information for interested reader to be able to reconstruct the 2-vertex transformation matrix completely. 

It is important to note that correlations $\langle k_i|\rarrow\rangle$ and $\langle k_o|\rarrow\rangle$ of the exit vertex are very different from the correlations obtained with 1-vertex statistics.
 It can be written using usual one-vertex statistics as:  $\langle k_i|\rarrow\rangle=\frac{\langle k_ik_o \rangle}{\langle k_o \rangle}$, while $\langle k_o|\rarrow\rangle=\frac{\langle k_o^2 \rangle}{\langle k_o \rangle}$.
Similarly, the correlations of in-vertex written by means of one-vertex statistics are $\langle q_i|\rarrow\rangle=\frac{\langle q_i^2 \rangle}{\langle k_o \rangle}$, while $\langle q_o|\rarrow\rangle=\frac{\langle q_iq_o \rangle}{\langle k_o \rangle}$. 

It can be shown that the correlations arising from transformation process are different from those that we would expect from the non-correlated network.
 The comparison between real correlations in the network and the ones expected from the configuration model is shown in Figure \ref{CorrVsRandCorr.eps}.

 Up to now we have shown that degree correlations can be strongly influenced by addition of bidirectional arcs. 
If the initial network is already very correlated transformation process tends to amplify these correlations compared to configuration model.
An example of such strongly correlated networks is the Barab\'{a}si-Albert directed network that we used for comparison \cite{Usmjereni BA}.
In this model new vertices are attached to the old ones proportionally to the sum of in--degree of the old vertices and some parameter $a$. 
In our case the parameter $a=1$ was chosen in the simulations and the starting network for BA evolution was Gilbert network of $10^3$ vertices connected with probability $0.01$. 
Other values of parameter $a$ were tested as well. 
It is known that the properties of the directed BA network do not depend on size or degree sequence of initial network in the thermodynamical limit. 
The final size of the simulated networks was $10^6$.

A very important assumption of this analysis is that the initial network does not mutate/evolve in any other way during the transformation process. 
However, reality is that many complex networks evolve during the course of time and are highly nonequilibrium systems~\cite{N03}. 
There is a myriad of different rules one can think of in order to simulate some aspect of network growth and to study the influence of the addition of the bidirectional arcs in all these cases is impossible. 
We decided to study how addition of reciprocal arcs changes some very well studied growth process. 
An obvious candidate for the study was preferential attachment growth with the addition of reciprocal arcs during the evolution process.

\section{Growth model}\label{ChapterPrefRecip}

\indent In this model the crucial idea is that at formation of directed arc between new and old vertex there is a transfer of information about that event from the pointing vertex to the pointed vertex. In that case the old vertex can return the newly formed arc to the new vertex, thus forming a bidirectional arc. In the model this process of information return will be modeled by the probability $r$ that the old vertex points back to the new vertex.

The model can be described as a variant of the directed network growth by means of preferential attachment and formation of reciprocal arcs. 
More precisely: in every time step $t$ there are $t$ vertices labeled from $0,...,t-1$ present in the network, and a new vertex, labeled with $t$  attaches to the network with $m$ outgoing arcs. 
Every of those $m$ arcs is attached to some already present vertex $s$ with probability proportional to the in--degree of the old vertex i.e. $P(t\rarrow s)\simeq\frac{k_i(s)}{\langle k_i \rangle t}$. 
If the network is grown only using this rule, the model is a variant of the BA model for the growth of directed network with the atractivness parameter $a=0$. 
The additional rule is that every of $m$ new arcs, with probability $r$, can recieve a reciprocal arc from the old vertex.
 With this additional rule the model is completely described. 
It is useful to note that based on previous work \cite{BA99b,DMS00} one can expect that for the value of parameter $r=0$ the network will have in--degree distribution with exponent $\gamma=-2$, and for value of parameter $r=1$ the network will have in--degree distribution with exponent $\gamma=-3$.

\indent Although we will later calculate the analytical expression for the joint degree probability distribution for general parameter $m$, we will first present the solution for $m=1$ because it is easier to write it in a closed form.
 We will use master equation to calculate joint degree probability distribution.
Let $p(k_i,k_o,s,t)$ present the probability that vertex introduced to the network at the moment $s$, at time $t$ possesses in--degree $k_i$ and out--degree $k_o$.
 In this treatment, for simplicity we will not use bidirectional degree because it would unnecessarily complicate the calculation.
 As the initial condition at time $t=1$ we choose the network of two reciprocally connected vertices  $s=0$ i $s=1$.
\begin{eqnarray}
\label{InitKondPreRecm1}
p(k_i,k_o,0,1)&=&\delta_{k_i,1}\delta_{k_o,1},\nonumber\\
p(k_i,k_o,1,1)&=&\delta_{k_i,1}\delta_{k_o,1}.
\end{eqnarray}

The probability $p(k_i,k_o,s,t)$ for $k_o\geq1$, $k_i > 0$, and $s<t$ is
\begin{eqnarray}\label{MicroMasterPreRecm1}
p(k_i,k_o,s,t)&=&\frac{k_i-1}{L_{in}(t)}(1-r)p(k_i-1,k_o,s,t-1)+\nonumber\\
&&\frac{k_i-1}{L_{in}(t)}rp(k_i-1,k_o-1,s,t-1)+\nonumber\\
&&\left(1-\frac{k_i}{L_{in}(t)}\right)p(k_i,k_o,s,t-1).
\end{eqnarray}
The function $L_{in}(t)$ is a random variable which is equal to the sum of all degrees present in the network at time $t$, 
\begin{eqnarray}\label{L(t)m1}
L_{in}(t)&=&\sum_{s=0}^{t-1} k_i(s)\nonumber\\
&=&\sum_{k_i} k_i\sum_{s,k_o}p(k_i,k_o,s,t-1).
\end{eqnarray}
The following approximation for the $L_{in}(t)$ is very reasonable for a very big network
\beq\label{aproxL(t)m1}
L_{in}(t)\simeq\langle L_{in}(t)\rangle=(1+r)t,
\eeq
i.e. we assume that the random variable $L_{in}(t)$ is well described by its expected value. 

The equation for the vertex $t$ which is just attaching to the network at the time $t$ is
\begin{equation}\label{RubniPreRecm1}
 p(k_i,k_o,t,t)=r\delta_{k_i,1}\delta_{k_o,1}+(1-r)\delta_{k_i,0}\delta_{k_o,1}.
\end{equation}
Probability that the vertex $s$ does not have any ingoing arc is
\beq\label{BezInPReREC}
p(0,k_o,s,t)=(1-r)\delta_{k_o,1}.
\eeq

 We sum the obtained joint probabilities $p(k_i,k_o,s,t)$ that the vertex $s$ at time $t$ has degrees $k_i$ and $k_o$, over all present vertices $s$ to get the probability $P(k_i,k_o,t)$, that the randomly chosen vertex at time $t$ has degrees $k_i$ and $k_o$ i.e. $P(k_i,k_o,t)=\sum_{s=0}^tp(k_i,k_o,s,t)/(t+1)$.
 We also assume that the distribution will be stable for large $t$ i.e. $P(k_i,k_o,t)\stackrel {t\rarrow\infty}{\longrightarrow}P(k_i,k_o)$. The described procedure results with the equations
\begin{equation}\label{Prvam1}
P(0,k_o)=(1-r)\delta_{k_o,1},
\end{equation}
\begin{equation}\label{Drugam1}
P(k_i\geq 1,1)=\frac{r(1+r)\delta_{k_i,1}+(k_i-1)(1-r)P(k_i-1,1)}{1+r+k_i},
\end{equation}
and
\begin{eqnarray}\label{Trecam1}
P(k_i\geq 1,k_o>1)&=&\frac{k_i-1}{1+r+k_i}[(1-r)P(k_i-1,k_o)\nonumber\\
&&+rP(k_i-1,k_o-1)].
\end{eqnarray}

\begin{figure}[t]
\centering
\includegraphics*[width=0.45\textwidth]{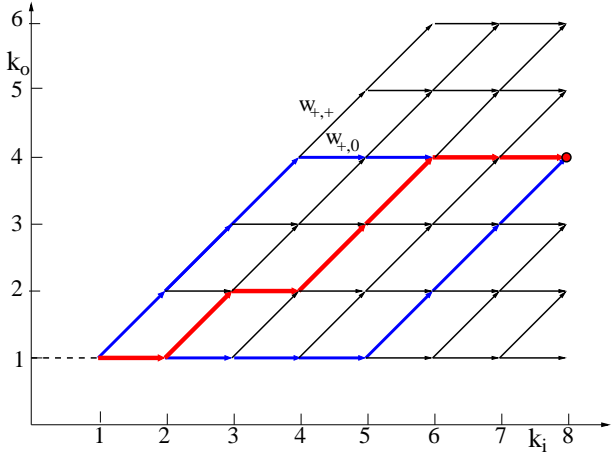}
\caption[Sumacija puteva za $m=1$] {\label{GridWalk} The event-set lattice used to calculate the stable distribution. The borders of the lattice segment in which paths contribute to the probability are designated with blue color.
 Red color designates one of the possible paths and arrows represent the possible directions of paths.}
\end{figure}
\begin{figure}[t]
\centering
\includegraphics*[width=0.5\textwidth]{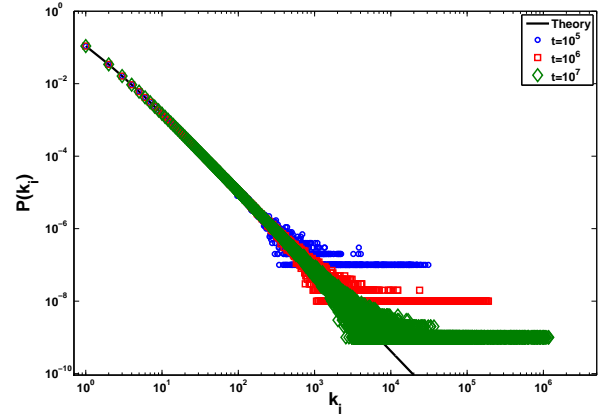}
\caption[Slaganje teorije i simulacija u modelu preferencijalnog rasta s o\v{c}uvanom recipro\v{c}nosti]{\label{TheoryAgainstSimulationPrefRecip} In this figure it can be seen the agreement between distribution of in--degrees of the process simulated over $100$ different realizations (markers) and the analytical solution (full line). The parameter $r$ has the value $0.2$. }
\end{figure}
\indent The Eq. (\ref{Drugam1}) shows that $P(1,1)=r(1+r)/(2+r)$. 
The simplest way to solve this set of equations is to sum contributions of all possible paths for probability distribution $P(k_i,k_o)$.
 Equations (\ref{Drugam1}) and  (\ref{Trecam1}) can be easily represented as the walk on the event lattice shown in Figure \ref{GridWalk}.
 The nodes of this lattice represent all the possible \emph{events} (degree combinations) of randomly choosing a vertex from the ensemble of networks generated by the studied process.
 Every movement to the right from the site $\{k_i-1,k_o\}$ to the site $\{k_i,k_o\}$ is multiplying the probability distribution attached to the site with the factor $w_{+,0}=(k_i-1)(1-r)/(1+r+k_i)$, while every diagonal movement from the site $\{k_i-1,k_o-1\}$ to the site $\{k_i,k_o\}$ represents multiplying the probability distribution with the factor $w_{+,+}=(k_i-1)r/(1+r+k_i)$.
 The value of the joint degree probability distribution $P(k_i,k_o)$ is therefore equal to sum of the contributions of all possible paths from site $\{1,1\}$ to the site $\{k_i,k_o\}$.
 Every path has $k_i-k_o$ movements to the right and $k_o-1$ diagonal movements and every of them has the same contribution
\begin{equation}\label{Doprinosm1}
r^{k_o-1}(1-r)^{k_i-k_o}P(1,1)\prod_{n=2}^{k_i}\frac{n-1}{1+r+n}.
\end{equation}
The number of distinct paths is $\binom{k_i-1}{k_o-1}$ which is equal to the number of combinations of factors $w_{+,0}$ and $w_{+,+}$. The general expression for the joint degree distribution is therefore equal to
\begin{widetext}
\begin{equation}\label{konacnom1}
P(k_i,k_o)=\Theta(k_i-k_o)\binom{k_i-1}{k_o-1}r^{k_o-1}(1-r)^{k_i-k_o}\frac{r(1+r)}{2+r}\frac{(k_i-1)!}{(r+3)_{k_i-1}},
\end{equation}
\end{widetext}
where denominator in the last factor $(r+3)_{k_i-1}$ represents Pochammer symbol, defined with relation $(x)_{n}=x(x+1)...(x+n-1)$.
 Nice property of this solution is that the correlations between degrees of one vertex are exactly computed and easily checked.
 In the limit of the big in--degree, using the representation of Pochammer symbol by means of Gamma functions $(x)_{n}=\frac{\Gamma(x+n)}{\Gamma(x)}$, it is easy to show that the asymptotic behavior of the last factor is $\lim_{k_i\rarrow\infty}\frac{(k_i-1)!}{(r+3)_{k_i-1}}=\Gamma(r+3)k_i^{-(2+r)}$.

\indent The exponent of the power-law-like tail is the property which can be easily checked for sufficiently large networks. We can sum over all the values of $k_o$ in the equations (\ref{Prvam1}), (\ref{Drugam1}), and (\ref{Trecam1}), and use the approximation of continuum to get the equation for the marginal in--degree distribution.
\begin{equation}\label{P(k_i)summ1}
(1+r)P(k_i)=(k_i-1)P(k_i-1)-k_iP(k_i).
\end{equation}
In the continuum approximation this equation can be written as:
\begin{equation}\label{ContPrefRecipm1}
 (1+r)P(k_i)=-\frac{d\left(k_iP(k_i)\right)}{dk_i},
\end{equation}
and has a solution
\begin{equation}\label{KontRjesP(k_i)m1}
 P(k_i)\sim k_i^{-(2+r)}.
\end{equation}

To check analytical solutions we have performed a series of simulations for different values of parameter $r$ and different network sizes. For all monitored parameters and sizes of network we found a nice agreement between analytical solution and simulations (Fig. \ref{TheoryAgainstSimulationPrefRecip}). 

\section{Growth model for a general parameter $m$}

\begin{figure}[t]
\centering
\includegraphics*[width=0.5\textwidth]{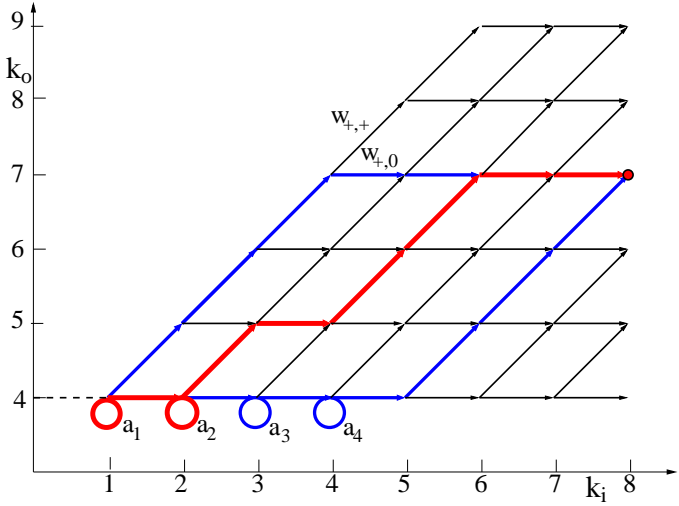}
\caption{\label{GridWalkm} The lattice used to calculate the stable distribution of the model for the parameter $m=4$. Blue color designates the borders of the lattice area within which the paths contribute to the joint probability $P(k_i=7,k_o=8)$, while the red path represents one of the possible contributing paths. Arrows represent the allowed directions of movement on the lattice, while the loops on the sites represent the additional coefficients $a_l$ contributing to joint degree probability on these sites.}
\end{figure}

\begin{figure}[t]
\centering
\includegraphics*[width=0.5\textwidth]{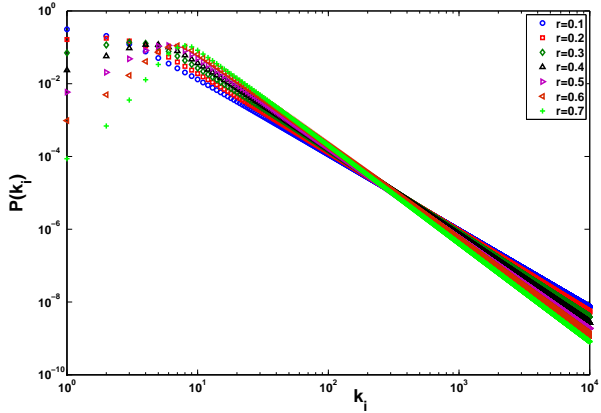}
\caption{\label{DEpendentR} The marginal distribution of the in--degree calculated from the theory for $m=10$ and different values of parameter $r$. The tail of distribution clearly depends on the the parameter $r$ and the power law character of the tail is easily checked. The shape and the position of the mode also depends on the value of $r$.}
\end{figure}

\begin{figure}[t]
\centering
\includegraphics*[width=0.5\textwidth]{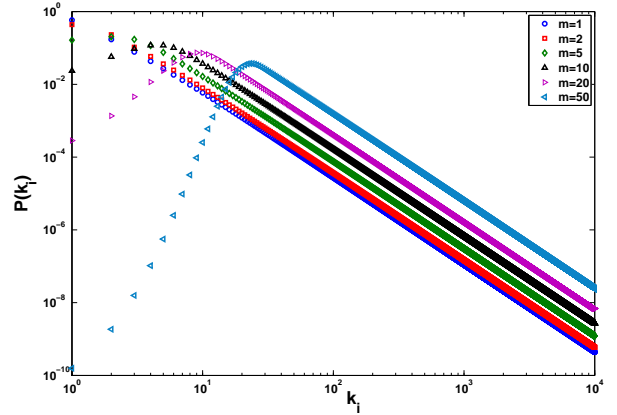}
\caption{\label{DEpendentM} The marginal distribution of the in--degree calculated from the theory for $r=0.4$ and different values of parameter $m$. The existence and the position of the mode strongly depend on the value of parameter $m$. The tail of distribution is independent of the parameter $m$ and the power law character of the tail is easily checked. }
\end{figure}

It is possible to calculate joint degree distribution of the model for the general parameter $m$. We again use master equation:
\begin{widetext}
\begin{eqnarray}\hspace{-0.2cm}
\label{MicroPrefRec}
p(k_i,k_o,s,t)&=&\sum_{l=0}^{m}\binom{m}{l}\left(\frac{k_i-l}{L_{in}(t)}\right)^l\left(1-\frac{k_i-l}{L_{in}(t)}\right)^{m-l}\Theta(k_i-l)\nonumber\\
&&\cdot\sum_{n=0}^{l}\binom{l}{n}r^n(1-r)^{l-n}p(k_i-l,k_o-n,s,t-1)\Theta(k_o-m-n),\nonumber\\
\end{eqnarray}
\end{widetext}
where $\Theta(x)$ represents usual Heaviside Theta function with convention $\Theta(0)=1$. Indices $m$, $l$ and $n$ combine all the possible combinations of: number of outgoing arcs formed on the new vertex, number of new arcs attached to the old vertex and the number of formed reciprocal arcs from which it is possible to create vertex with given degrees.
 To ease the calculation, we allowed the formation of multiple arcs between two vertices, which in the thermodynamical limit does not influence the exact solution.
 The boundary condition of this set of equations is
\begin{equation}\label{RubniPrefRecip}
p(k_i,k_o,t,t)=\sum_{n=0}^{m}\binom{m}{n}r^n(1-r)^{m-n}\delta_{k_i,n}\delta_{k_o,m}.
\end{equation}
We again sum over all vertices and approximate the function $L_{in}(t)$ with its expected value $L_{in}(t)\simeq\langle L_{in}(t)\rangle=(1+r)mt$.
Using equation (\ref{RubniPrefRecip}) and assuming stable degree distribution in the thermodynamical limit the following equation for the joint degrees distribution is obtained
\begin{widetext}
\begin{eqnarray}\label{JointPrefRecip}
\frac{1+r+k_i}{1+r}P(k_i,k_o)&=&\sum_{n=0}^{m}\binom{m}{n}r^n(1-r)^{m-n}\delta_{k_i,n}\delta_{k_o,m}+\frac{k_i-1}{1+r}\Theta(k_i-1)\Theta(k_o-m-1)rP(k_i-1,k_o-1)\nonumber\\
&&+\frac{k_i-1}{1+r}\Theta(k_i-1)\Theta(k_o-m)(1-r)P(k_i-1,k_o).\nonumber\\
\end{eqnarray}
\end{widetext}
This equation can be solved in a manner similar to equations (\ref{Prvam1}), (\ref{Drugam1}), and (\ref{Trecam1}). 
In Figure \ref{GridWalkm} we present the event-set lattice with the contributions of the paths to the joint probability attached to every site.
The contribution of every path is again identical for every lattice bond as soon as the path detaches from the line $k_o=m$.
 The total contribution of the paths differ only in the number of steps made on line $k_o=m$ for the $1<k_i\leq m$.
For $1<k_i\leq m$ and $k_o=m$ the joint degree probability is
\begin{eqnarray}\label{kilessthenm}
P(k_i,m)&=&\frac{1+r}{1+r+k_i}\sum_{l=1}^{k_i-1}a_l(1-r)^{k_i-l}\prod_{j=l}^{k_i-1}\frac{j}{1+r+j}\nonumber\\
&&+\frac{1+r}{1+r+k_i}a_{k_i},
 \end{eqnarray}
and $a_l$ represents the probability of binomial distribution $a_l=\binom{m}{l}r^l(1-r)^{m-l}$. The equation for the case $k_i>m$ and $k_o=m$ is
\begin{equation}\label{kigeqm}
 P(k_i,m)=(1-r)^{k_i-m}P(m,m)\prod_{j=m+1}^{k_i}\frac{j-1}{1+r+j}.
\end{equation}
The contribution of paths which separated from the line $k_o=m$ i $1<k_i\leq m$ at the site $(k_i=k_i',k_o=m)$ is:
\begin{widetext}
\begin{equation}\label{doprinosoodvajanju}
P(k_i,k_o)_{\pi(k_i',m)}=P(k_i',m)\binom{k_i+k_o-m-k_i'-2}{k_i-k_i'-1}(1-r)^{k_o-m}r^{k_i-k_i'-k_o+m}\frac{(k_i-1)!}{(k_i'-1)!}\frac{(r+3)_{k_i'-1}}{(r+3)_{k_i-1}}
 \end{equation}
\end{widetext}
 where $\pi(k_i',m)$ represents the sum over the possible paths after detachement from the line $k_o=m$. The whole solution is now easy to write using equations (\ref{kilessthenm}), (\ref{kigeqm}), and (\ref{doprinosoodvajanju}), but it is complicated and not very informative.
Nevertheless, it is interesting to monitor the behavior of Eq. (\ref{kilessthenm}).
 It can be verified that for $P(k_i<m,k_o)$ the joint degree distribution can increase depending on the initial parameters of the model.
 On the other hand, in the limit $k_i\gg m$ we expect the fall of the joint probability distribution. 
This implies that for certain range of parameters the distribution has a nontrivial mode.
 Such a behavior can be easily checked if the equation (\ref{JointPrefRecip}) is summed over all out--degrees $k_o\in[m,\infty\rangle$, to obtain the marginal distribution of in--degree.
 In the range $k_i\in[1,m]$ the solution is 
\begin{equation}
\label{InPrefRecipm2}
P(k_i)=\frac{1+r}{1+r+k_i}\left(a_{k_i}+\sum_{l=1}^{k_i-1}a_l\frac{(k_i-1)!}{(l-1)!}\prod_{j=l}^{k_i-1}\frac{1}{1+r+j}\right).
\end{equation}
\begin{figure}[t]
\centering
\includegraphics*[width=0.5\textwidth]{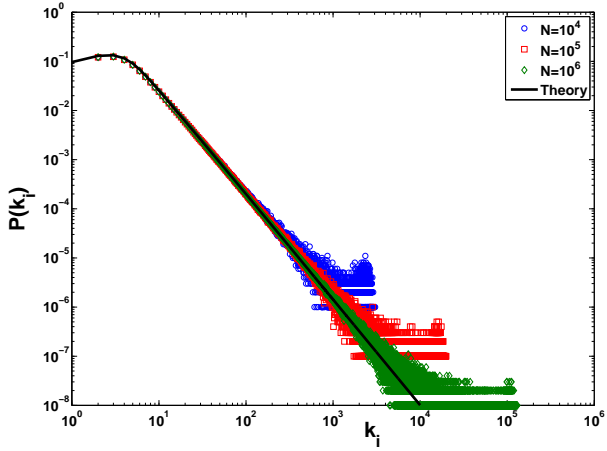}
\caption{\label{TheoVsSimGenm} The marginal distribution of the in--degree calculated from theory (full line) and simulations (markers), for $3$ different sizes of final networks. The averaging was performed over $100$ simulation realizations, and parameters used are $r=0.15$ and $m=18$. There is an excellent agreement between theory and simulations.}
\end{figure} 
The marginal in--degree distribution obtained analytically coincide with the simulations rather well as shown on Fig. \ref{TheoVsSimGenm}. 
The modal character of the in--degree distribution is easily observed in this equation. 
The dependence of mode on the parameters $r$ and $m$ is shown in Figures \ref{DEpendentR} and \ref{DEpendentM}. 

The other important property of the distribution is its power law behavior in the tail.
 As can be seen in Figure \ref{DEpendentM} the exponent of the tail does not depend on the parameter $m$. Power law behavior of the tail is governed only by the parameter $r$ as shown in Figure \ref{DEpendentR}. Indeed in the continuum approximation, valid for $k_i\gg 1$, the equation for the in--degree marginal distribution is
\begin{equation}
 \label{ContMPrefRecip}
(1+r)P(k_i)\sim-\frac{d\left(k_iP(k_i)\right)}{d k_i}.
\end{equation}
The solution of this equation is
\begin{equation}
 \label{ContResult}
P(k_i)\sim k_i^{-(2+r)},
\end{equation}
and dependence of the value of the power law exponent with respect to parameter $r$ is very clear. The equation for the power law exponent $\gamma=-2-r$ also confirms our claim that the model behavior interpolates between usual BA model with the exponent $-3$ and the directed BA model with the exponent $-2$.
\begin{figure}[t]
\centering
\includegraphics*[width=0.5\textwidth]{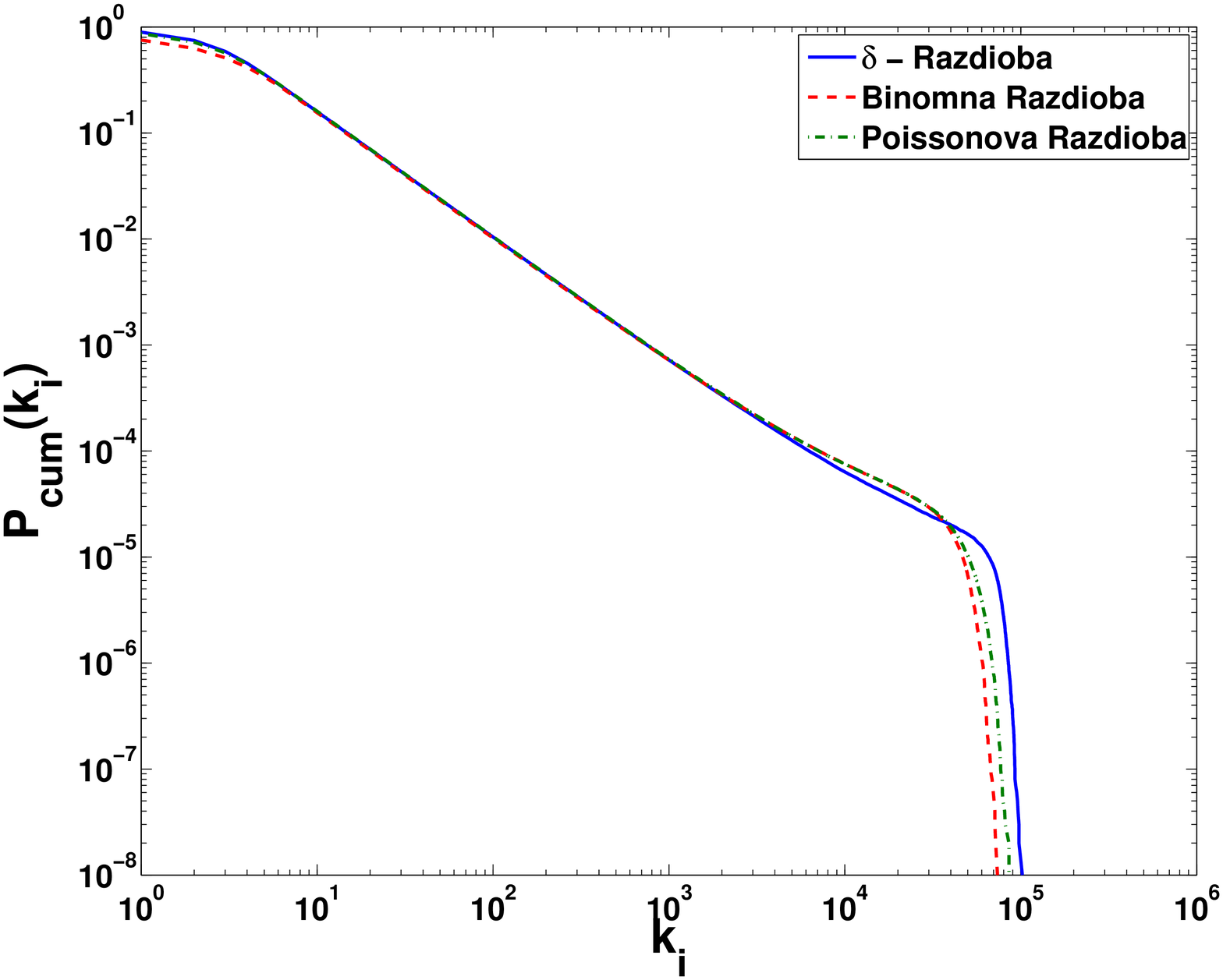}
\caption[Razdioba ulaznih stupnjeva za razne razdiobe parametra $m$]{\label{StartingDistros}This figure shows that there is no important difference between behavior of the in--degree distribution for the different choices of the vertex $t$ out-degree distribution at time $t$. The simulations are presented for the parameters  $m=10$, $r=0.2$, $N=10^6$ and $100$ realizations. The presented distributions are cumulative in--degree distributions which are more comparable in the case of simulated models. The broader distributions of initial out--degree show a bit smaller maximal degree, which we explain by stronger competition for the new vertices.}
\end{figure}

It is important to mention that this analytical discussion is valid up to certain point also for a little bit broader class of growth models. In the analytic treatment the distribution of the outgoing arcs of the new vertex $t$ (\ref{RubniPrefRecip}) at the time $t$ is a delta function. We can expect that this reasoning can be applied for a more general class of distributions for the outgoing arcs of the vertex $t$ at time $t$ with the assumption that the mean field approximation is valid. In particular, we expect that this consideration will be valid for all unimodal discrete distributions with fast decaying tails. To test this assumption we examined cases in which the out--degree of the vertex $t$ at time $t$ is drawn from the Binomial and Poisson distribution.

 The Poisson distribution  
\begin{equation}
 \label{PoissRecip}
P(k_o|m)=\frac{m^{k_o}e^{-m}}{k_o!},
\end{equation}
is determined only by parameter $m$.
 For every monitored $m$ of the original model, we made a new set of simulations with Poisson distribution with the same $m$.
 For Binomial distribution
\begin{equation}
 \label{BinomRecip}
P(k_o|m,Z)=\binom{Z}{k_o}\left(\frac{m}{Z}\right)^{k_o}\left(1-\frac{m}{Z}\right)^{Z-k_o},
\end{equation}
the case is a little bit more complicated because it is defined with two parameters: $m$ - the expected number of outgoing arcs and $Z$ maximal allowed out-degree of the vertex $t$ at time $t$. For a broad choice of values of parameter $Z$, the results were very similar to the ones expected from the original model as $N\gg Z$ as can be seen in Fig. \ref{StartingDistros}.

\section{Conclusion} 
\indent
We have shown that reciprocal arcs can significantly influence the degree correlations in complex networks.
 In the first part of the paper we laid down a way to investigate influence of randomly distributed bidirectional arcs on the overall degree correlations and have shown how this hypothesis can be tested in the case of real networks. 
We also studied a simple model of the network growth which conserves expected fraction of reciprocal arcs. 

The analysis laid out in the first part of this paper focuses on the degree correlations represented as average product moments.
 This analysis has its positive and negative side.
 The average product moments clearly do not contain as much information as the average neighbor degree functions~\cite{N03} and this is an obvious shortcoming of such a measure.
 On the other hand it is exactly the reason why product moments can be very useful for case studies.
 In the case of very correlated networks, sometimes the frequency of degree statistics for the large degrees is so scarce that it effectively shrinks the available configuration space for the null models which are trying to preserve correlations found in the network.
 This reduction of available configuration space can sometimes be so huge that for connected pairs of vertices with large degrees any result different from the already observed in the network is almost impossible to realize.
 In this case the product moments incorporate in themselves much larger number of viable network realizations so that the analysis of network with correlated null models is much better founded.

In the companion paper~\cite{RecipWiki}, we apply the theory presented in this paper to show that Wikipedia networks cannot be explained by random distribution of bidirectional arcs on the static network. In the same paper we used the presented growth model to explain in--degree distribution of the Wikipedia networks with very good results.

From the theoretical point of view this model helps to understand possible mechanisms which create modes in the degree distributions of different scale-free directed networks.
 Furthermore this model is a good candidate to explain other empirical directed networks which combine power-law tails and nontrivial mode of the degree distribution and the future work in this direction is clearly needed.
 It also represents one of the simplest growth models which preserves some type of local correlations. 
 It is our opinion that the understanding of interrelations between different types of correlations in complex networks heavily depends on such growth models.
The validation of this claim is an important task for our future research. 

\section{Acknowledgments}

This work was financed by the Ministry of Education, Science and Sports of the Republic of Croatia under the contract No. 098-0352828-2863 and by the INFN, Italy. Vinko Zlati\'c is thankfull for support of G. Caldarelli and would like to thank A. Gabrielli for reading the manuscript.

\end{document}